# Immersive Video Effects Persist Beyond Exposure:

## From Psychological Closeness to Self-Initiated Behaviour in a Two-Week Comparison with 2D Video

**TOIDA Koichi[1], FURUYA Shinnosuke[2], MIURA Shimpei[2], KOBAYASHI Yuki[1], and MEGURO Shingo[2]**

[1] MESON, Inc.
[2] Hakuhodo DY Holdings Inc., Marketing Technology Center.

Corresponding author: TOIDA Koichi (e-mail: koichi.toida@meson.tokyo)

**ABSTRACT** Immersive Video (IV), a live-action stereoscopic 180° video format, has been shown to elicit a strong sense of Presence and enhance psychological closeness to individuals depicted in the video. However, whether these effects persist beyond exposure and are accompanied by subsequent self-initiated behaviour remains unclear. In this study, 84 participants were assigned to either an IV or a 2D video group, and psychological and behavioural outcomes were tracked over a two-week period. Psychological outcomes were assessed before and immediately after exposure and at approximately 3, 7, and 14 days post-exposure, while self-initiated behaviour was assessed during follow-up. Psychological closeness remained significantly higher in the IV group than in the 2D group for up to 7 days and remained elevated above baseline within the IV group at 14 days. Lasting-impression ratings were also significantly higher in the IV group throughout the follow-up period. Among participants who completed both scheduled follow-up assessments, 45.9% in the IV group met the operational criterion for early self-initiated engagement, compared with 13.2% in the 2D group. Furthermore, a higher proportion of participants in the IV group reported active or socially oriented behaviours. These findings suggest that IV can elicit psychological effects that persist beyond exposure and may encourage subsequent self-initiated engagement.

## 1 INTRODUCTION

The extent to which psychological effects induced by media experiences persist after exposure has ended is relevant to both memory research and media psychology. Classical memory research has documented a decline in the availability of encoded information over time, as illustrated by Ebbinghaus's forgetting curve (Ebbinghaus, 1885/1913). A modern replication reproduced a similar pattern of forgetting over intervals ranging from 20 minutes to 31 days (Murre & Dros, 2015). However, what persists over time is not limited to the information itself. Emotionally arousing events tend to form more enduring memories than less arousing events, a phenomenon associated with interactions between emotional arousal and memory consolidation processes (McGaugh, 2004; LaBar & Cabeza, 2006). Therefore, when examining the persistence of effects induced by a media experience, it is necessary to distinguish cognitive retention, as reflected in the accurate recall of information, from the persistence of affective and psychological effects arising from the experience.

The relationship between media format and persistence has long been investigated through comparisons of audio and audiovisual media. In a comparison of learning through radio and television, no clear medium-specific differences were found in the amount of knowledge acquired or in subsequent forgetting. However, emotional impressions elicited by the two media differed, and these differences remained relatively stable over time (Helmreich, 1976). This finding suggests that retention of learned information and persistence of experiential impressions can follow different temporal patterns. Adding visual information does not necessarily slow the forgetting of learned material, even when medium-specific psychological impressions persist over time.

Evidence on learning and retention in immersive VR is mixed. HMD viewing improved spatial-memory recall over desktop viewing (Krokos et al., 2019), while a meta-analysis found modest learning benefits that could persist (Wu et al., 2020). Conversely, HMD-based visuomotor adaptation was associated with greater cognitive load and poorer 24-hour retention and context transfer (Juliano et al., 2022). Immersion alone does not guarantee better retention.

Psychosocial effects may also persist. Embodied perspective-taking can influence self–other merging, attitudes, and helping behaviour (Ahn et al., 2013). In an eight-week follow-up study using first-person VR to simulate homelessness, VR produced longer-lasting positive attitudes than traditional methods, although self-reported empathy did not differ significantly. VR participants were also more likely to sign a petition supporting people experiencing homelessness, but this behavioural outcome was measured immediately after exposure rather than during follow-up (Herrera et al., 2018). Thus, effects may vary by outcome and assessment time.

Another question explored in VR research is the extent to which immersive experiences can substitute for real-world experiences. In tourism, for example, the potential use of VR as an alternative to physical visits has been discussed (Guttentag, 2010), and comparisons of 360° VR-based virtual visits with actual on-site visits have suggested the two can elicit similar experiences in terms of Presence—the subjective sense of “being there” in a mediated environment (Slater, 2009)—and engagement (Wagler & Hanus, 2018). Live performances offer a suitable context for examining such relationships between on-site and mediated experiences. The same performance can be experienced by attending it in person or by viewing it through conventional 2D video or immersive video, and direct comparisons among these formats have been conducted (Scorolli et al., 2023).

In research on live performances, live music has been shown to elicit stronger and more dynamic activity in emotion-related brain regions than recorded music (Trost et al., 2024). Furthermore, when the same performance was experienced either live or on a screen from the same viewing position in the same theatre, greater subjective pleasure and arousal, along with stronger responses on some physiological measures, were observed in the live condition (Gabriel et al., 2025). Beyond the immediate experience, a large-scale natural experiment found that concert attendance increased subsequent online listening to the performing artist. Increased listening was also observed among socially connected non-attendees, although this spillover effect was found only for popular artists (Ternovski & Yasseri, 2020). Together, these findings suggest that live performances can elicit strong emotional responses during the event and can be followed by changes in subsequent listening behaviour.

Within this context, Immersive Video (IV) differs from both conventional 2D video and computer-generated virtual environments. By combining the visual realism of live-action footage with stereoscopic 180° video, HMD-based presentation, head-motion-contingent changes in viewing direction, and spatial audio, IV may allow a recorded event to be experienced not merely as video viewed on an external screen but as an event unfolding within the space occupied by the viewer. Although viewers cannot navigate or physically interact with the recorded environment, head rotation allows them to select the viewing direction even when no explicit virtual body or avatar is presented. When the viewing position is experienced as their own location, a sense of Presence may arise. Thus, IV can elicit a sense of "being there" through viewpoint-based self-location, even without active manipulation of the environment (Toida, 2026a). This experiential structure is consistent with meta-analytic evidence that greater technological immersion tends to enhance Presence (Cummings & Bailenson, 2016). Among the components of Presence, Place Illusion refers to the subjective experience of being located within the presented space (Slater, 2009), whereas Social Presence refers to the feeling that another person encountered through a medium is present as a social other in the same space. The latter has been shown to increase with visual, spatial, and interactive cues (Oh et al., 2018).

Toida et al. (2026) investigated the psychosocial effects of IV by manipulating filming distance. A live performance by a Japanese idol group was recorded as stereoscopic 180° live-action IV, with the camera-to-performer distance varied between 1.2 m and 7.6 m. The near-distance condition increased several Presence-related measures, including Possible Actions, Social Presence, and Observability. Psychological overlap between the self and the performers depicted in the video was also assessed using the Inclusion of Other in the Self (IOS) Scale, and a greater increase in psychological closeness was observed in the near-distance condition than in the far-distance condition. These findings indicate that filming distance in IV is not merely a variable governing framing or composition but a design parameter capable of modulating the viewer's psychological relationship with the target.

Building directly on this earlier work, the present study compares the effects of the IV stimulus previously presented in the near-distance condition (Toida et al., 2026) with those of conventional 2D video and extends the assessment into the post-exposure period. Here, psychological closeness refers not merely to favourability or interest but to the extent to which the target is perceived as psychologically close to the self. Because the previous study focused primarily on immediate post-exposure responses, it remained unclear whether the increase in psychological closeness observed after IV exposure would persist over time or be accompanied by subsequent self-initiated behaviour.

An important distinction in evaluating the effects of immersive media is between the magnitude of an immediate psychological response and its persistence over time. A larger initial effect does not necessarily imply a slower subsequent decline, and an advantage over another medium is not equivalent to a sustained increase above baseline. These distinctions are particularly relevant when examining psychological closeness, which may remain elevated even after the initial viewing experience has ended. Furthermore, the persistence of psychological effects does not necessarily lead to subsequent behaviour. Although behavioural intention is associated with action, changes in intention do not always translate into corresponding behavioural changes (Ajzen, 1991; Webb & Sheeran, 2006). Assessing self-initiated activities alongside psychological outcomes therefore provides a complementary perspective on the longer-term consequences of immersive experiences, without assuming a direct causal relationship between psychological closeness and subsequent behaviour.

Accordingly, the present study extends the findings of Toida et al. (2026) along three dimensions: comparison across media formats, persistence over time, and self-initiated behaviour following exposure. In this study, persistence is operationally defined as the maintenance of media-induced psychological changes beyond exposure, rather than memory consolidation in its narrower sense. Psychological closeness was the primary outcome, while favourability, interest, behavioural intention, and lasting impression were assessed as secondary outcomes. Self-initiated behaviour related to the performers was also assessed during the follow-up period.

Three hypotheses were formulated. First, near-distance IV, which elicited relatively high levels of Presence in the previous study, was expected to produce a greater immediate increase in psychological closeness than conventional 2D video (H1). Second, the advantage of IV over 2D video in psychological closeness was expected to persist throughout the follow-up period rather than being confined to the immediate post-exposure assessment (H2). Third, compared with 2D video, IV was expected to yield a higher proportion of participants meeting the operational criterion for self-initiated engagement during the early follow-up period, as well as a higher proportion engaging in active or socially oriented behaviours during follow-up (H3).

# 2 METHOD

## 2.1 Overview

This study employed a two-week longitudinal experimental design, with media condition (IV vs. 2D) as a between-participants factor and measurement time point as a within-participants factor. At T0, assessments were conducted before (Pre) and immediately after (Post) video exposure, followed by assessments approximately 3 days (T1), 7 days (T2), and 14 days (T3) after exposure. The permitted response windows were 2–4 days for T1, 6–8 days for T2, and 13–15 days for T3.

## 2.2 Sample Size Planning

Sample-size planning was informed by data from a separate one-month follow-up assessment of participants in the previous experiment reported by Toida et al. (2026). In that follow-up, a small-to-moderate within-participant effect ($dz$ = 0.36) was observed for the change in psychological closeness from Post to one month after exposure. This value was used as a reference for planning rather than as a direct estimate of the between-group effect in the present study. A standardised between-group effect size of $d$ = .35–.40 was assumed for the IV–2D comparison. For a two-sided independent-samples comparison with $\alpha = .05$ and power of $1 - \beta = .80$, approximately 130 participants per group would be required for $d = .35$, compared with 100 participants per group for $d = .40$.

However, practical constraints associated with in-person HMD sessions limited recruitment to 84 participants, with 42 assigned to each condition. The longitudinal analyses used linear mixed-effects models (LMMs) to account for repeated measurements and baseline scores. A planned-missingness design was also adopted to reduce participant burden and the potential influence of repeated assessments, although this necessarily reduced the number of observations available at each follow-up time point. These analytical and design choices did not eliminate the limitations associated with the achieved sample size.

## 2.3 Participants

Prospective participants were recruited through a panel maintained by a research agency. Recruitment was not restricted to existing fans of the idol group featured in the stimulus. However, to reduce the likelihood of including individuals with a strong aversion to idol-related content, participants were required to have had a favourite idol or singer, either currently or in the past. Additional eligibility criteria included visual acuity of at least 1.0 with unaided vision or contact lenses and no use of medication that could induce drowsiness. Individuals employed in advertising- or XR-related professions were excluded. Given the potential discomfort associated with viewing performers at close range, social anxiety, gaze anxiety, and interpersonal distance were also assessed before participation, with reference to previous research (Nagata et al., 2013; Langer et al., 2014; Iachini et al., 2014). Candidates were also asked to imagine a person appearing approximately 1 m away and rate how strongly they would wish to move away on a five-point scale. Those selecting the highest response category were excluded. These pre-participation assessments were not included as covariates in the reported analyses.

Allowing participants to choose between attending an in-person experimental session in the IV condition and participating online in the 2D condition could have confounded media condition with willingness to attend, time commitment, and compensation. To avoid participant self-selection, eligible candidates from a panel maintained by the research agency who were able to participate in either format were randomly assigned to the IV or 2D condition by the agency.

A total of 84 participants were recruited, with 42 assigned to the IV group and 42 to the 2D group. Recruitment and allocation were organised to achieve comparable age and gender distributions across the two groups. Specifically, approximately five participants from each combination of age group (20s, 30s, 40s, or 50s) and gender were assigned to each group.

The study was conducted in accordance with the third edition of the Ethical Principles of Psychologists issued by the Japanese Psychological Association (2011), following an internal review and approval of the research protocol by the affiliated organisation. Participants were informed in advance of the study objectives, procedures, voluntary nature of participation, and right to withdraw at any time. Written informed consent was obtained from participants in the IV group, whereas informed consent was obtained online from those in the 2D group. As part of the informed consent procedure, participants agreed not to disclose information obtained through the study. Participants in the IV group were also informed that they could discontinue the experiment immediately at their own discretion if symptoms such as cybersickness occurred.

## 2.4 Stimuli and Presentation Apparatus

The experimental stimulus consisted of a live performance of “Shukkō” by 12 fourth-generation trainee members of the Japanese idol group STU48. The IV condition used the stereoscopic 180° live-action IV stimulus presented in the near-distance condition of Toida et al. (2026). The camera-to-performer distance was approximately 1.2 m, a condition previously shown to elicit higher levels of Presence and a greater increase in psychological closeness than the far-distance condition of 7.6 m.

The IV stimulus was encoded at a resolution of 4,320 × 4,320 pixels per eye, 90 fps, and 150 Mbps, and was presented using AVPlayer on the Apple Vision Pro (M2; Apple). Audio was presented using Ambisonics-based 3DoF spatial audio. These specifications were identical to those of the near-distance stimulus used in the previous experiment. Further details of the filming conditions and implementation requirements are provided in Toida et al. (2026).

For the 2D condition, a 1080p, 16:9 monoscopic video derived from the image for one eye of the same stereoscopic recording was used. Thus, the performers, music, performance, and temporal sequence were identical across conditions. The conditions differed in stereoscopy, field of view, head-motion-contingent changes in viewing direction, spatial audio presentation, display device, viewing environment, and the resulting spatial experience. In particular, IV was presented using an HMD in a controlled experimental setting, whereas 2D video was viewed online using participants' own devices. The video lasted approximately 3 minutes and 40 seconds in both conditions.

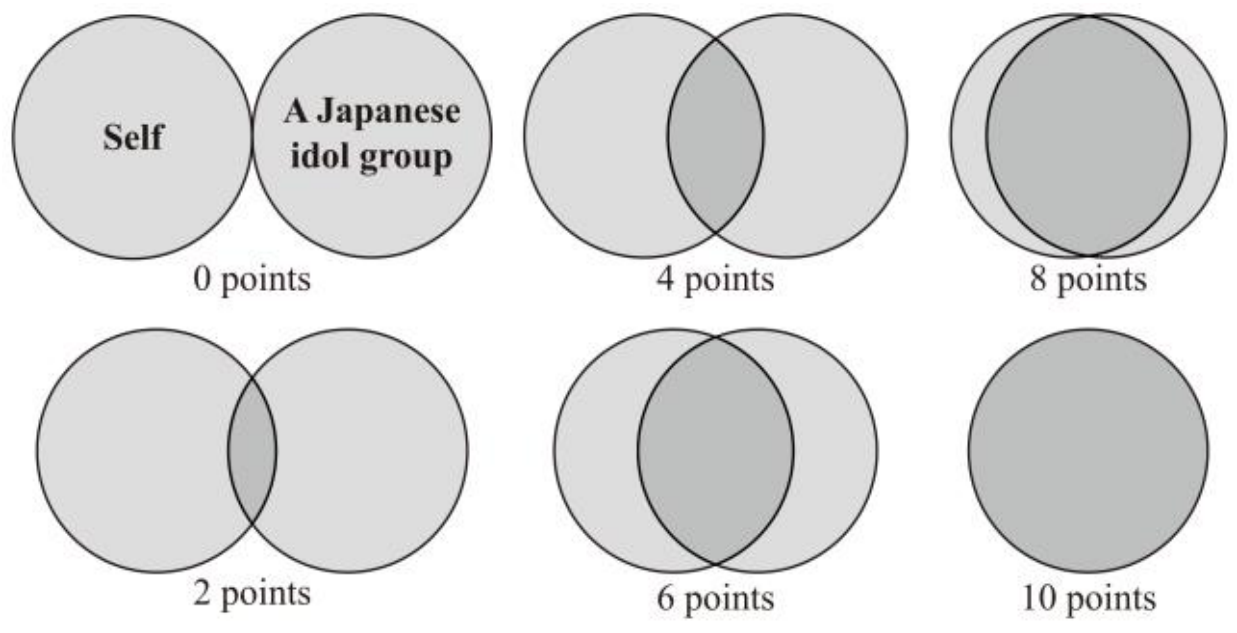

**FIGURE 1** 11-point IOS scale used to measure psychological closeness

## 2.5 Procedure

The T0 experimental session for the IV group was conducted in a quiet experimental setting. The procedure consisted of participant check-in, an explanation of the study and informed consent, the Pre questionnaire, instructions on wearing and operating the Apple Vision Pro, video exposure, and the Post questionnaire, in that order. Participants completed the Pre questionnaire on an iPad and then experienced the IV stimulus while wearing the Apple Vision Pro. After viewing the video, they removed the HMD and completed the Post questionnaire on the iPad. The entire session lasted approximately 30 minutes per participant.

The T0 session for the 2D group was conducted online. Each participant was provided with a unique URL through the research agency and accessed the study using their own computer, smartphone, or tablet. The procedure consisted of informed consent, presentation of precautions and study instructions, entry of the participant ID, verification of the viewing environment, the Pre questionnaire, video exposure, and the Post questionnaire, in that order. The video was delivered via streaming, and access through each URL was restricted to a single viewing. Participants were required to complete the T0 session within a 24-hour response window. The seek bar was hidden, and participants could not proceed to the Post questionnaire until the video had finished playing. Before viewing, participants were instructed to ensure a stable internet connection and a quiet environment, use earphones, adjust screen brightness to a comfortable level, and watch the video in landscape orientation when using a smartphone. In addition, a short sample video was played before the experimental stimulus to allow participants to verify that both video and audio were functioning properly.

For the follow-up assessments at T1–T3, participants were not required to complete all three assessments. Instead, they were assigned to one of three follow-up schedules (T1–T2, T1–T3, or T2–T3), limiting each participant to two follow-up assessments. Accordingly, a balanced planned-missingness design was adopted, with 28 participants per condition assigned to each follow-up time point. Planned-missingness designs have been proposed as a means of reducing participant burden while maintaining sufficient information at the group level across measurement occasions (Graham et al., 2006).

The decision to limit follow-up assessments to two per participant was motivated by considerations beyond response burden. In memory research, the act of retrieving or recalling a previous experience can itself enhance subsequent retention, a phenomenon known as the testing effect (Roediger & Karpicke, 2006). Furthermore, merely asking questions about behavioural intentions has been shown to influence subsequent behaviour, an effect known as the mere-measurement effect (Morwitz et al., 1993; Godin et al., 2008). Because the present study aimed to assess the persistence of the initial video experience and subsequent self-initiated behaviour, the number of follow-up assessments was limited to reduce the possibility that repeated measurements would reactivate the initial experience or influence subsequent behaviour.

## 2.6 Measures

The primary outcome measure was psychological closeness, assessed using the Inclusion of Other in the Self (IOS) Scale. The original IOS Scale (Aron et al., 1992) is a single-item measure consisting of seven pairs of circles representing the self and another person, ranging from two separate circles to substantially overlapping circles. Respondents select the pair that best represents their current relationship. The IOS Scale has demonstrated high convergent validity with several measures of relationship closeness (Gächter et al., 2015). To capture more fine-grained changes, the present study adopted an 11-point configuration based on the extended IOS Scale described by Baader et al. (2024), presenting 11 levels of overlap between the circles. Responses were scored from 0, indicating no overlap, to 10, indicating maximum overlap. Participants were asked to indicate their perceived psychological closeness to the target group by selecting the degree of overlap between the two circles that best represented their relationship with the group. Higher scores indicated greater psychological closeness (Figure 1).

Favourability was assessed by asking participants how favourably they viewed the target group, interest by asking how much they wanted to know more about the group, and behavioural intention by asking how much they wanted to watch videos or listen to music by the group. All three measures used seven-point rating scales. The interest measure was developed with reference to the four-phase model of interest development, which proposes that situational interest triggered by external stimuli can be maintained and further developed through renewed engagement (Hidi & Renninger, 2006). Although behavioural intention is a proximal predictor of actual behaviour (Ajzen, 1991), meta-analytic evidence indicates that experimentally induced changes in behavioural intention produce smaller changes in subsequent behaviour (Webb & Sheeran, 2006). Accordingly, behavioural intention was treated as distinct from self-initiated behaviour in the present study.

Lasting impression was assessed from Post onwards using the question "To what extent does the impression of the video you watched during the initial session still remain with you?", with responses recorded on a seven-point rating scale. This measure did not assess the accuracy of recalled or recognised video content. Rather, it was intended to capture the subjective persistence of the initial viewing experience. Accordingly, lasting impression was treated as distinct from objective memory retention.

At T0 Pre, participants were also asked about their prior awareness of the target idol group, frequency of contact with the group during the preceding month, and previous engagement in related behaviours. At each participant's second follow-up assessment, additional exploratory measures, including spatial vividness and the subjective sense of re-experiencing, were collected. However, these measures were not included in the hypothesis tests reported in this paper.

## 2.7 Classification Using the 5A Framework

At T1–T3, participants were asked whether they had encountered the target group or related content following the initial video exposure and whether such contact had been incidental or voluntary. Participants reporting voluntary contact were asked to identify the specific activities they had undertaken and the cumulative number of days on which they had voluntarily engaged with the target or related content since the initial exposure. Each calendar day was counted once, regardless of the number of contacts on that day. Reported activities were examined to identify qualifying self-initiated behaviour.

The behavioural analyses were restricted to participants who completed both of their scheduled follow-up assessments, comprising 37 participants in the IV group and 38 in the 2D group. These denominators were used consistently for the cumulative behavioural comparisons and differed from the numbers of participants who provided psychological assessments at any individual follow-up time point.

For participants assessed at T1, early self-initiated engagement was identified from their T1 responses. For participants whose first scheduled follow-up was T2, early engagement was classified retrospectively when they reported voluntary contact on four or more days by T2. This rule provided an operational estimate of early engagement but did not establish the exact date on which the first voluntary activity occurred. Participants were retained in the denominator regardless of their assigned follow-up schedule.

For cumulative analyses, information from both scheduled follow-up assessments was combined for each participant. Once a qualifying behaviour had been identified, the corresponding attainment status was retained in subsequent cumulative classifications. The resulting classifications represented whether qualifying activities had been reported by each observation point, rather than independently observed behavioural events at every time point.

Specific activities were classified using the 5A framework proposed by Kotler et al. (2016). Table 1 presents the operational definitions and classification criteria. The framework was used as a descriptive taxonomy rather than a psychological stage model, and the categories were not assumed to be mutually exclusive or to represent a required sequence of behavioural progression.

## 2.8 Statistical Analysis

Because the follow-up assessments at T1–T3 included missing observations by design, conventional repeated-measures analysis of variance (ANOVA), which requires observations at every time point for all participants, was not applied. Psychological closeness, favourability, interest, and behavioural intention were analysed using linear mixed-effects models (Laird & Ware, 1982). All available observations at Post, T1, T2, and T3 from the 83 participants with valid T0 data were included, yielding 241 observations for each measure. A participant-specific random intercept was included to account for within-participant repeated measurements, and the corresponding Pre score was entered as a covariate.

**TABLE 1** Operationalization of the 5A stages and classification criteria

| 5A stage | Operational criterion | Behaviours classified in this study |
|---|---|---|
| **A1 Aware** | Prior awareness of the target group | Already knew the target group at baseline (T0) |
| **A2 Appeal** | Voluntary engagement indicating continued interest | Voluntarily viewed the target group's social-media posts or online videos; listened to or watched songs, music videos, or live-performance videos; newly followed or subscribed to official/member accounts; engaged with posts (e.g., liked, saved, commented on, or shared them). |
| **A3 Ask** | Active information seeking | Searched for the group, members, or songs; viewed official websites, profiles, appearance information, or live-event information; searched for information about live events, fan clubs, CDs, streaming, or merchandise. |
| **A4 Act** | Concrete participation or transaction | Purchased, registered for, subscribed to, or participated in live events, fan clubs, CDs, streaming services, merchandise, or related activities. |
| **A5 Advocate** | Social-media interaction or interpersonal advocacy-related behaviour | Liked, saved, commented on, or shared social-media posts; talked about or recommended the target group to family members, friends, or acquaintances. |

Note. The questionnaire combined liking, saving, commenting, and sharing social-media posts into a single response option. This composite item was eligible for classification under both A2 and A5, following the original coding scheme. Consequently, A5 attainment should not be interpreted as unambiguous evidence of interpersonal recommendation.

Let $Y_{it}$ denote the score for participant $i$ at time point $t$, $G_i$ denote the group indicator coded as 1 for IV and 0 for 2D, $T_{kit}$ denote dummy variables representing T1, T2, and T3, with Post as the reference time point, and $P_i$ denote the Pre score. The model was specified as follows:

$$Y_{it} = \beta_0 + \beta_G G_i + \sum_{k=1}^{3} \beta_{Tk} T_{kit} + \sum_{k=1}^{3} \beta_{GTk} (G_i T_{kit}) + \beta_P P_i + b_{0i} + \varepsilon_{it}$$

Fixed effects were estimated using maximum likelihood, and individual coefficients were evaluated using Wald $z$ tests. The group × time interaction was assessed using a three-degree-of-freedom Wald test encompassing the three interaction terms. IV–2D differences at each time point were estimated as linear contrasts of the fixed effects. To control the family-wise error rate, $p$ values for the four between-group comparisons were adjusted using the Holm method (Holm, 1979).

For psychological closeness, the primary outcome, supplementary analyses were conducted to determine whether the increase observed in the IV group remained elevated above Pre levels at each time point, independently of the between-group comparisons. Paired-samples $t$ tests comparing each post-exposure score with the corresponding Pre score were performed using participants who had provided valid responses at each time point. The resulting $p$ values were adjusted for four comparisons using the Holm method. These analyses were treated as supplementary assessments of persistence, separate from the primary LMM analysis.

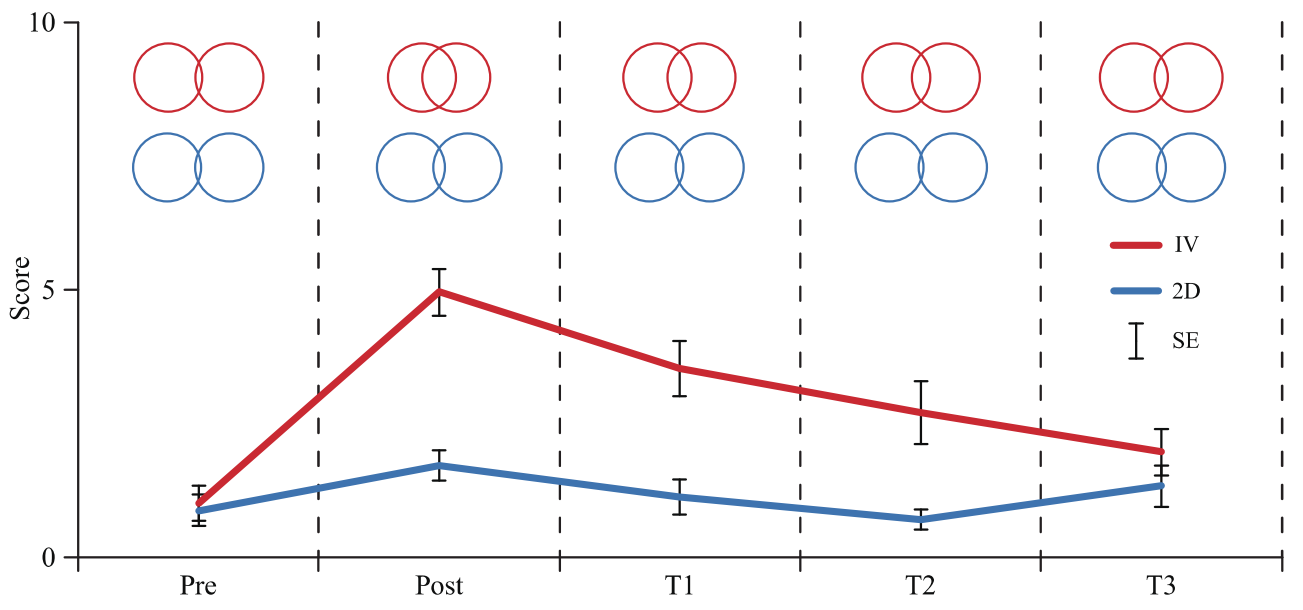

**FIGURE 2** Average Psychological Closeness (IOS Score) at Each Time Point

Because lasting impression was not assessed at Pre, the following model was applied:

$$LastingImpression \sim Group \times Time + (1 \mid Participant)$$

The proportion of participants meeting the operational criterion for early self-initiated engagement was compared between groups using a two-sample test of proportions, with the 37 IV and 38 2D participants who completed both scheduled follow-up assessments as denominators. Relative risk and its 95% confidence interval were also calculated. Participants who reported at least one behaviour classified as A3, A4, or A5 during the follow-up period were categorised as having reached "A3 or higher". This composite category was used to identify participants who engaged in active information seeking or socially oriented behaviour; it did not imply sequential progression through the 5A stages. The proportions of participants meeting this criterion were compared between the IV and 2D groups. Fisher's exact test was used for A5 because of the small expected cell counts. For attainment of A3 or higher, a two-sample test of proportions was used as the primary analysis, supplemented by Fisher's exact test as a sensitivity analysis. The significance level was set at a two-sided $\alpha = .05$ for all analyses.

All analyses were conducted using Python and R.

# 3 RESULTS

## 3.1 Participant Flow and Attrition

At T0, 42 participants took part in each of the IV and 2D conditions. One participant in the 2D group was excluded because their T0 response was submitted outside the 24-hour response window. Consequently, valid T0 data were available from 42 participants in the IV group and 41 in the 2D group.

Of these participants, 37 in the IV group and 38 in the 2D group completed both of their scheduled follow-up assessments, corresponding to completion proportions of 88.1% and 92.7%, respectively. The remaining participants did not complete both scheduled assessments, although some contributed outcome data at individual follow-up time points. Because the planned-missingness design assigned participants to different follow-up schedules, absence of a response at a particular time point did not necessarily indicate attrition.

The longitudinal analyses of psychological outcomes included all available score observations from the 83 participants with valid T0 data, yielding 241 post-exposure observations across Post, T1, T2, and T3. Thus, the numbers of observations in the psychological analyses differed from the numbers of participants completing both scheduled follow-up assessments. The behavioural analyses, in contrast, were restricted to the 37 IV and 38 2D participants who completed both scheduled assessments.

At baseline, prior awareness of the target group was reported by 31 of 42 participants in the IV group and 24 of 41 in the 2D group. Among those who completed both follow-up assessments, the corresponding numbers were 26 of 37 and 22 of 38, respectively. These descriptive differences in prior awareness were considered when interpreting the behavioural findings.

## 3.2 Psychological Closeness

Descriptive statistics for psychological closeness are presented in Figure 2 and Table 2. In the IV group, the mean IOS score increased markedly from Pre ($M = 0.95$, $SE = 0.30$) to Post ($M = 4.88$, $SE = 0.42$), followed by a gradual decline at T1 ($M = 3.39$, $SE = 0.51$), T2 ($M = 2.50$, $SE = 0.53$), and T3 ($M = 2.12$, $SE = 0.44$). In contrast, the corresponding scores in the 2D group were $M = 0.90$ ($SE = 0.28$) at Pre, $M = 1.71$ ($SE = 0.27$) at Post, $M = 1.19$ ($SE = 0.31$) at T1, $M = 0.70$ ($SE = 0.19$) at T2, and $M = 1.27$ ($SE = 0.39$) at T3, indicating a substantially smaller descriptive increase from Pre to Post.

The LMM adjusted for Pre scores revealed a significant group × time interaction (Wald $\chi^2_{(3)} = 18.66$, $p < .001$), indicating that the temporal trajectories of psychological closeness differed between the IV and 2D groups. The fixed effects are presented in Table 3. At Post, the adjusted difference between the IV and 2D groups was $\beta = 3.160$ ($SE = 0.459$, $z = 6.88$, $p < .001$). The interaction coefficients were negative at T1 ($\beta = -1.176$, $p = .027$), T2 ($\beta = -1.421$, $p = .008$), and T3 ($\beta = -2.195$, $p < .001$), indicating that the initially large advantage of IV at Post attenuated over time. The Pre score was also a significant predictor ($\beta = 0.271$, $p = .008$). The random-intercept variance was 2.217, the residual variance was 2.154, and the intraclass correlation coefficient (ICC) was .507.

The baseline-adjusted IV–2D difference was 3.160 points at Post ($SE = 0.459$, 95% CI [2.260, 4.060], Holm-adjusted $p < .001$), 1.984 points at T1 ($SE = 0.533$, 95% CI [0.939, 3.029], Holm-adjusted $p < .001$), and 1.739 points at T2 ($SE = 0.537$, 95% CI [0.687, 2.791], Holm-adjusted $p = .002$). At T3, the estimated difference had attenuated to 0.965 points ($SE = 0.544$, 95% CI [−0.101, 2.031], Holm-adjusted $p = .076$). Although the confidence interval included zero and the between-group contrast was no longer statistically significant, this result does not establish that the difference had disappeared. The baseline-adjusted advantage of IV over 2D video remained statistically significant through approximately 7 days after exposure.

Supplementary within-group analyses compared psychological closeness at each post-exposure time point with the corresponding participants' own Pre scores. In the IV group, the mean increase at Post was $M_{diff} = 3.93$, $t_{(41)} = 8.61$, Holm-adjusted $p < .001$, $d_z = 1.33$; at T1, $M_{diff} = 2.25$, $t_{(27)} = 3.77$, Holm-adjusted $p = .002$, $d_z = 0.71$; and at T2, $M_{diff} = 1.42$, $t_{(25)} = 2.90$, Holm-adjusted $p = .011$, $d_z = 0.57$. Even at T3, psychological closeness remained significantly elevated above baseline ($M_{diff} = 1.44$, $t_{(24)} = 3.07$, Holm-adjusted $p = .011$, $d_z = 0.61$). In contrast, none of the four comparisons with Pre reached statistical significance in the 2D group after Holm adjustment. These findings indicate two temporal patterns: a statistically significant advantage of IV over 2D video in psychological closeness through approximately one week, and a significant elevation above baseline within the IV group at two weeks.

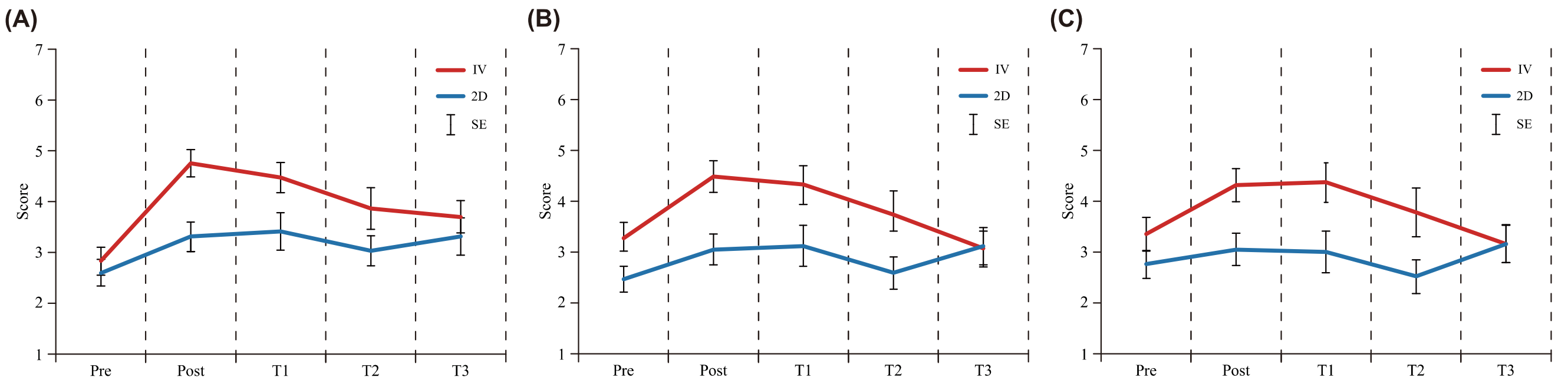

**FIGURE 3** Changes in Average Scores across Time Points for the Secondary Psychological Measures: **(A)** Favourability **(B)** Interest **(C)** Behavioural Intention

## 3.3 Secondary Psychological Measures

The same baseline-adjusted LMM structure was applied to favourability, interest, and behavioural intention. The adjusted IV–2D differences at each time point are presented in Table 4. Significant group × time interactions were observed for favourability ($\chi^2_{(3)} = 12.67$, $p = .005$), interest ($\chi^2_{(3)} = 17.55$, $p < .001$), and behavioural intention ($\chi^2_{(3)} = 9.16$, $p = .027$), indicating that post-exposure trajectories differed between groups for all three measures (Figure 3).

Lasting-impression scores were higher in the IV group than in the 2D group at every post-exposure assessment (Figure 4). The group × time interaction was not statistically significant ($\chi^2_{(3)} = 6.58$, $p = .087$), whereas the model-estimated IV–2D contrasts were statistically significant at each time point: +1.973 at Post (Holm-adjusted $p < .001$), +1.791 at T1 ($p < .001$), +1.398 at T2 ($p = .001$), and +1.106 at T3 ($p = .007$). All reported $p$ values for the time-point contrasts were adjusted using the Holm method.

## 3.4 Self-Initiated Behaviour

Using the operational criterion for early self-initiated engagement, 17 of 37 participants (45.9%) in the IV group and 5 of 38 participants (13.2%) in the 2D group were classified as meeting the operational criterion for early self-initiated engagement described in Section 2.7. A two-sample test of proportions revealed a significant between-group difference ($z = 3.12$, $p = .0018$), with a relative risk (*RR*) of 3.49 (95% CI [1.44, 8.49]).

Cumulative classification of post-exposure behaviours according to the 5A framework showed that 9 participants in the IV group and 7 in the 2D group had reached A2 (Appeal) by the end of follow-up. The corresponding numbers were 16 and 10 for A3 (Ask), none in either group for A4 (Act), and 8 and 2 for A5 (Advocate). Figure 5 illustrates the cumulative behavioural classifications based on the available follow-up responses.

Because the behavioural categories were not mutually exclusive, the composite category "A3 or higher" was also examined at the participant level. During follow-up, 18 of 37 participants (48.6%) in the IV group and 10 of 38 participants (26.3%) in the 2D group reported at least one behaviour classified as A3, A4, or A5. The difference was 22.3 percentage points ($RR = 1.85$). Although the two-sample test of proportions indicated a statistically significant difference ($z = 2.00$, $p = .046$), Fisher's exact test did not ($p = .058$). Thus, the evidence for a between-group difference in this composite outcome was sensitive to the choice of test.

**TABLE 2** Descriptive statistics for psychological closeness (IOS)

| Time point | Immersive Video | | 2D | |
|---|---|---|---|---|
| | *n* | *M ± SE* | *n* | *M ± SE* |
| Pre | 42 | 0.95 ± 0.30 | 41 | 0.90 ± 0.28 |
| Post | 42 | 4.88 ± 0.42 | 41 | 1.71 ± 0.27 |
| T1 | 28 | 3.39 ± 0.51 | 26 | 1.19 ± 0.31 |
| T2 | 26 | 2.50 ± 0.53 | 27 | 0.70 ± 0.19 |
| T3 | 25 | 2.12 ± 0.44 | 26 | 1.27 ± 0.39 |

**TABLE 3** Fixed effects of the linear mixed model for psychological closeness

| Fixed effects | *β* | *SE* | *z* | *p* | 95% CI |
|---|---|---|---|---|---|
| Intercept | 1.463 | 0.339 | 4.31 | <.001 | [0.798, 2.128] |
| IV | 3.16 | 0.459 | 6.88 | <.001 | [2.260, 4.060] |
| T1 | −0.687 | 0.38 | −1.80 | 0.071 | [−1.432, 0.059] |
| T2 | −0.710 | 0.375 | −1.89 | 0.058 | [−1.446, 0.025] |
| T3 | −0.588 | 0.38 | −1.55 | 0.122 | [−1.334, 0.157] |
| IV × T1 | −1.176 | 0.53 | −2.22 | 0.027 | [−2.215, −0.137] |
| IV × T2 | −1.421 | 0.534 | −2.66 | 0.008 | [−2.467, −0.375] |
| IV × T3 | −2.195 | 0.541 | −4.06 | <.001 | [−3.256, −1.135] |
| Pre | 0.271 | 0.102 | 2.65 | 0.008 | [0.070, 0.471] |

**TABLE 4** Adjusted IV–2D differences in psychological measures

| Measure | Post | T1 | T2 | T3 |
|---|---|---|---|---|
| Psychological Closeness | +3.160<br>$p < .001$ | +1.984<br>$p < .001$ | +1.739<br>$p = .002$ | +0.965<br>$p = .076$ |
| Favourability | +1.239<br>$p < .001$ | +0.802<br>$p = .017$ | +0.702<br>$p = .032$ | +0.404<br>$p = .171$ |
| Interest | +0.955<br>$p = .005$ | +0.485<br>$p = .281$ | +0.609<br>$p = .195$ | −0.211<br>$p = .529$ |
| Behavioural Intention | +0.948<br>$p = .018$ | +0.800<br>$p = .082$ | +0.793<br>$p = .082$ | +0.109<br>$p = .766$ |

Note. *p* values were adjusted using the Holm method for the four time-point comparisons

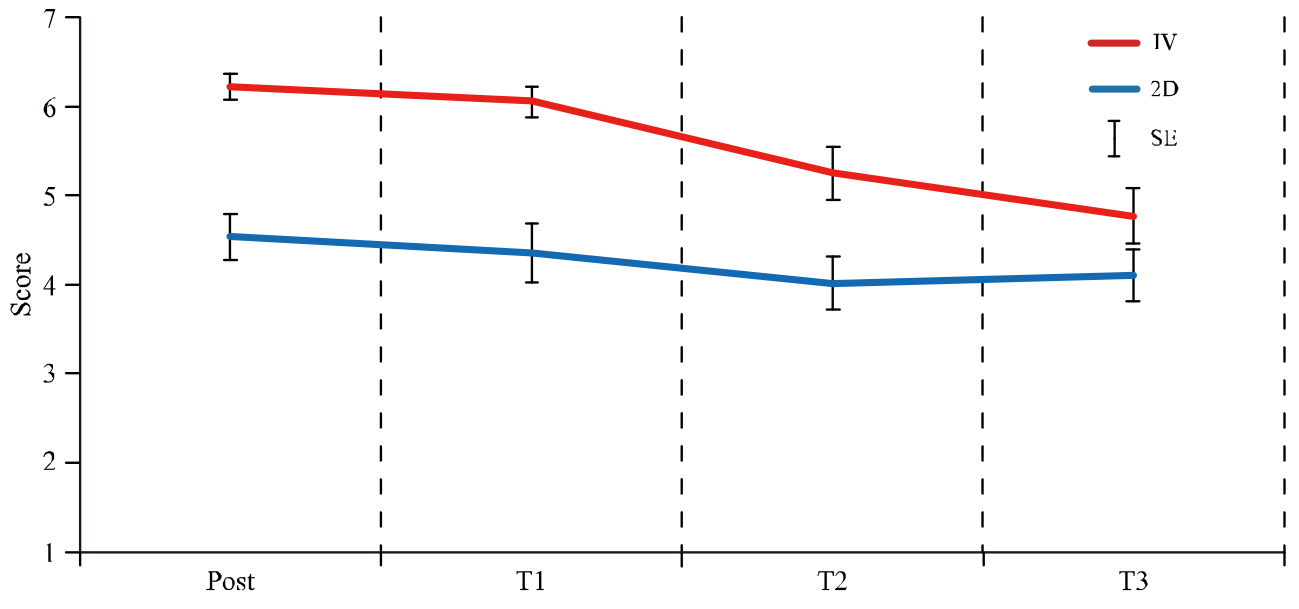

**FIGURE 4** Average Scores at Each Time Point: Lasting Impression

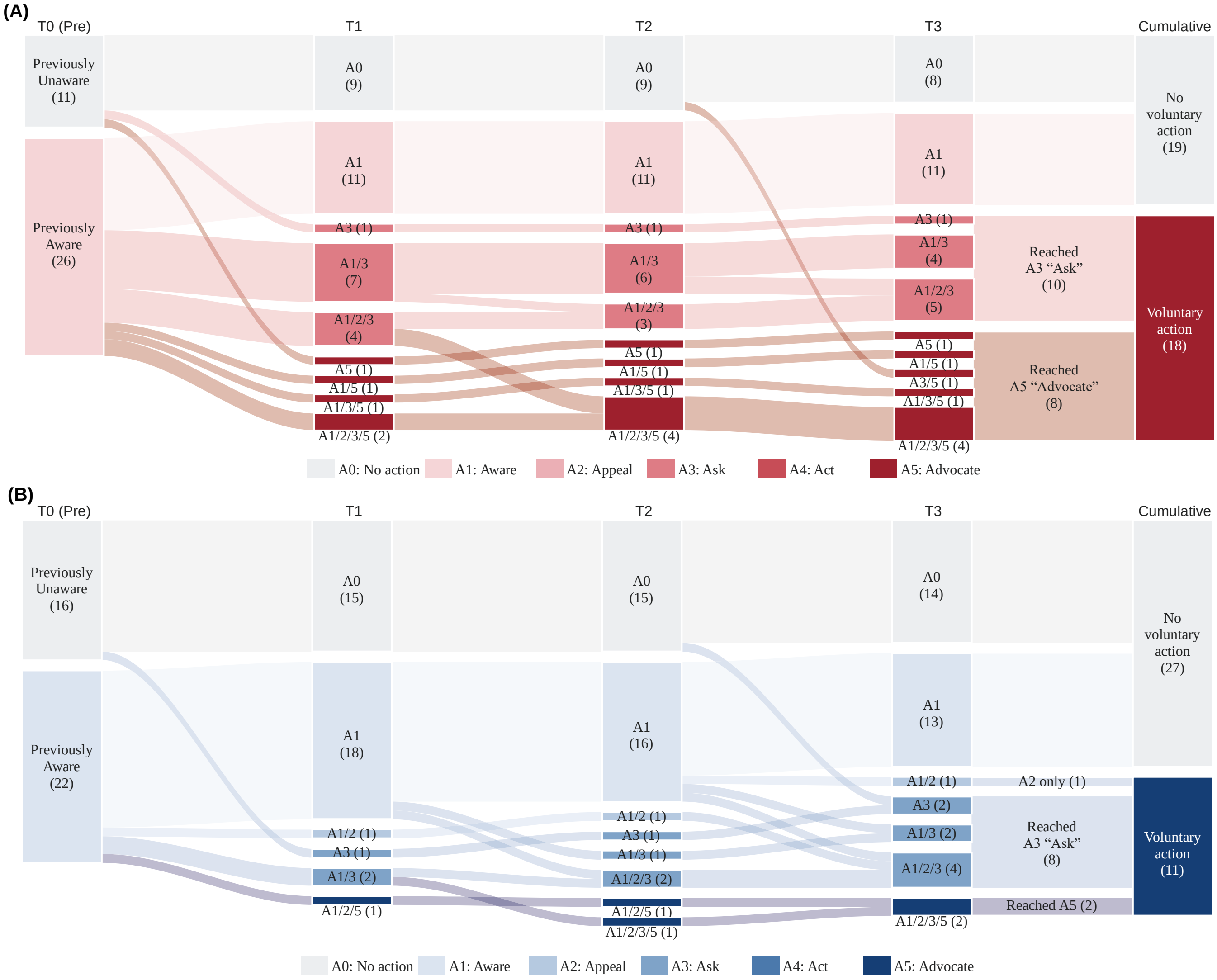

**FIGURE 5** Cumulative 5A behavioural classifications across follow-up assessments: **(A)** IV group; **(B)** 2D group

Note. The figure includes participants who completed both scheduled follow-up assessments (IV: $n = 37$; 2D: $n = 38$). Cumulative states were constructed from available responses using the predefined coding rules and do not necessarily represent directly observed behaviour at every displayed time point. The terminal flow groupings are mutually exclusive visual summaries, whereas the category-specific A2–A5 attainment counts reported in the Results are based on potentially overlapping activity classifications.

For A5 specifically, qualifying behaviour was identified in 8 of 37 participants (21.6%) in the IV group and 2 of 38 participants (5.3%) in the 2D group (Fisher's exact $p = .047$). As defined in Table 1, this category included a composite social-media interaction item as well as interpersonal recommendation-related activities. Therefore, A5 attainment did not exclusively indicate direct recommendations to others. No participants in either group reported purchases, registrations, or event participation classified as A4.

# 4 DISCUSSION

Three hypotheses were examined in this study. H1 was supported, as psychological closeness was significantly higher in the IV group than in the 2D group immediately after exposure. H2 was partially supported: the advantage of IV persisted through approximately seven days, but the between-group difference was no longer statistically significant at 14 days. H3 received mixed support. A significantly higher proportion of participants in the IV group met the operational criterion for early self-initiated engagement, whereas the statistical significance of the between-group difference in active or socially oriented behaviours depended on the test employed.

## 4.1 Persistence of Psychological Closeness

The principal finding of this study is that the increase in psychological closeness following IV exposure did not dissipate immediately, with a statistically significant advantage over 2D video persisting through approximately seven days. Furthermore, IOS scores in the IV group remained significantly elevated above baseline at 14 days. Toida et al. (2026) previously reported that near-distance IV elicited higher levels of Presence and a greater immediate increase in psychological closeness than far-distance IV. The present study extends these findings by showing that increased psychological closeness following IV exposure can remain detectable beyond the initial viewing session.

The IOS Scale does not simply measure liking; rather, it assesses perceived overlap between the self and another person (Aron et al., 1992; Gächter et al., 2015). Therefore, the finding that IOS scores in the IV group remained above Pre levels at 14 days should not be interpreted merely as evidence that participants remembered the video. Instead, it indicates that the target group continued to be perceived as psychologically closer to the self than before exposure. Whether this sustained perception reflects a lasting change in self–other representation or other psychological processes remains to be determined.

This persistence may be associated with the distinctive spatial and social experience afforded by IV. From the perspective of Place Illusion, an immersive environment presented through an HMD can allow viewers to experience themselves as being located within the presented space rather than observing it from outside (Slater, 2009). Social Presence, meanwhile, refers to the extent to which another person encountered through a medium is experienced as present in the same space (Oh et al., 2018). Hall (1966) distinguished four zones of interpersonal distance—intimate, personal, social, and public—with personal distance extending approximately from 0.45 to 1.2 m. The stimulus used in the present study was filmed with the performers positioned approximately 1.2 m from the camera, near the outer boundary of personal distance. The previous experiment found that this near-distance condition elicited higher levels of Social Presence, Possible Actions, and Observability than the far-distance condition. These spatial and social cues may have contributed to a sense of psychological closeness to the performers.

Embodied experiences in immersive environments have been shown to alter self-other merging and influence subsequent attitudes and behaviours (Ahn et al., 2013). Similarly, immersive perspective-taking has been associated with longer-lasting changes in attitudes, alongside differences in behaviour measured immediately after exposure (Herrera et al., 2018). In light of these findings, the IV stimulus used in the present study may have temporarily altered participants' psychological representations of the performers, not merely by enlarging their images but by establishing spatial relationships with them within an egocentric reference frame. This may have produced a greater initial psychological change than 2D video, part of which persisted over time.

Taken together, these findings are consistent with the possibility that the spatial and social characteristics of IV contribute to an immediate increase in psychological closeness, part of which persists beyond exposure. However, the previous and present studies involved different participants, and Presence was not directly measured in the present study. Therefore, the proposed causal sequence linking Presence, psychological closeness, and subsequent persistence remains to be established.

## 4.2 Conceptualising Persistence

When discussing persistence in psychology, it is important to distinguish it from memory consolidation. In its narrower sense, consolidation refers to the time-dependent neural processes through which initially unstable memory representations become stabilised as enduring long-term memories (McGaugh, 2004). In classical research on forgetting, retention is commonly operationalised in terms of the extent to which previously learned information can subsequently be recalled or recognised (Murre & Dros, 2015). Because the present study did not assess the accuracy of participants' memory for performers' names, clothing, lyrics, or scenes, it cannot establish that IV enhanced objective memory consolidation relative to 2D video.

The present study instead examined the persistence of psychological closeness and the subjective impression of the initial experience. Specifically, it assessed whether participants continued to perceive the target group as psychologically close, whether the impression of the initial video remained, and whether participants subsequently engaged in self-initiated behaviour related to the performers. Previous research comparing television and radio has shown that medium-specific emotional impressions can persist over time even when differences in the forgetting of learned information are not observed (Helmreich, 1976). These findings suggest that the persistence of media-induced psychological effects should be distinguished from objective memory retention.

Neuroscientific evidence has also shown that emotionally arousing experiences are more likely to form enduring memories than less arousing experiences, partly through interactions within memory systems involving the amygdala and medial temporal lobe (McGaugh, 2004; LaBar & Cabeza, 2006). Although these mechanisms were not directly examined in the present study, multiple features of IV—including the close proximity of performers, gaze cues, live-action representations of human bodies, stereoscopic disparity, head-motion-contingent visual information, and spatial audio—may have increased the salience of the experience as an event personally experienced by the viewer.

The observed temporal patterns differed descriptively across measures. The IV–2D difference in psychological closeness remained statistically significant through T2 but not at T3, whereas lasting-impression ratings were significantly higher in the IV group at every post-exposure assessment. However, the group × time interaction for lasting impression was not statistically significant, and the two measures used different response scales. Therefore, these findings do not establish that psychological closeness and lasting impression attenuate at different rates. Nevertheless, the two measures capture distinct aspects of persistence—the maintenance of a psychological relationship with the target and the lingering impression of the experience itself—and may provide complementary insights into the persistence of IV-induced psychological effects.

Furthermore, from T2 to T3, mean scores for psychological closeness, favourability, interest, and behavioural intention increased in the 2D group but declined in the IV group. Because the participants assessed at these time points were not identical under the planned-missingness design, these descriptive trends do not necessarily represent within-participant changes. Previous research has shown that repeated questioning can influence subsequent attitude evaluations (Sturgis et al., 2009) and that past experiences can be recalled more positively over time (Mitchell et al., 1997). One possible interpretation is that a common upward influence of recall and re-evaluation operated in both groups, while the IV group additionally experienced attenuation of its initially stronger psychological effects, resulting in an overall decline. This interpretation offers a perspective on psychological persistence as a dynamic phenomenon shaped by multiple overlapping processes.

## 4.3 Transition to Self-Initiated Behaviour

The second major finding of this study is that the between-group differences extended beyond psychological rating scales to self-reported voluntary behaviour following exposure. Although behavioural intention is an important proximal predictor of behaviour, a persistent gap exists between intention and actual behaviour. A meta-analysis of 47 experiments showed that experimentally induced moderate-to-large changes in behavioural intention produced substantially smaller changes in subsequent behaviour (Webb & Sheeran, 2006). In the present study, significant IV–2D differences in behavioural intention were limited to Post after Holm adjustment, whereas a significant between-group difference in self-initiated behaviour was observed during the early follow-up period. Self-reported voluntary engagement therefore provides more direct evidence of subsequent behaviour than a measure of behavioural intention alone.

Under the operational classification of early self-initiated engagement, 45.9% of participants in the IV group met the criterion, compared with 13.2% in the 2D group; the proportion was approximately 3.5 times as high in the IV group. This finding indicates that a higher proportion of participants in the IV group showed evidence of voluntary engagement during the early follow-up period. In research on interest development, situational interest triggered by external stimuli is considered capable of developing into maintained situational interest and eventually individual interest through renewed engagement (Hidi & Renninger, 2006). From this perspective, IV may have provided a strong initial stimulus for subsequent voluntary activities, including information seeking and repeated viewing.

The behavioural classifications further suggest that the effects of IV may extend beyond early self-initiated engagement to more active and socially oriented forms of involvement. During follow-up, 48.6% of participants in the IV group and 26.3% in the 2D group reported at least one behaviour classified as A3, A4, or A5. In particular, behaviours classified as A5 were reported by 8 participants in the IV group and 2 in the 2D group. Importantly, the reported behaviours concerned not the promotion or recommendation of the IV experience itself, but voluntary activities directed towards the performers, such as seeking information about them and interacting with related content on social media. These findings suggest that psychological changes elicited by IV may extend beyond the experimental setting and manifest as self-initiated engagement with the target.

No participants in either group reached A4 (Act) during the two-week follow-up period. Within the 5A framework, Act includes transactional behaviours such as purchasing or registering for a service, whereas Advocate includes recommending the target to others. The 5A customer path is not necessarily linear, and Advocate does not require prior attainment of Act (Kotler et al., 2016). Therefore, the absence of A4 behaviour does not establish that IV preferentially promoted non-transactional over transactional engagement. It indicates only that no qualifying A4 activity was reported within the observation period.

The observed behavioural pattern is qualitatively consistent with previous findings concerning live performances. Live music has been shown to elicit heightened emotional and physiological responses during performance (Trost et al., 2024; Gabriel et al., 2025). Separately, increased online listening following concert attendance, including spillover to socially connected non-attendees under certain conditions, has been documented (Ternovski & Yasseri, 2020). These findings suggest that the effects associated with live performances can extend beyond the event itself, although the processes underlying immediate emotional responses and subsequent engagement may differ.

IV is not equivalent to attending a live performance in person. In-person performances involve elements that cannot be fully reproduced through recorded video, including actual co-presence with performers, joint attention among audience members, crowd responses, and opportunities for real-time interaction. Nevertheless, the IV stimulus used in the present study was recorded from a camera position approximately 1.2 m from the performers and presented through stereoscopic video and spatial audio. The previous experiment showed that this near-distance stimulus elicited higher levels of Social Presence and Possible Actions than the far-distance condition. Compared with conventional 2D video, these characteristics may contribute to the subjective impression of being situated within a performance rather than viewing it from an external perspective. From a phenomenological perspective, Toida (2026b) argued that immersive media can allow recorded events to be experienced as unfolding in the present, distinguishing experiential presentness from objective simultaneity.

This sense of experiencing a recorded event as unfolding in the present may help explain why IV was associated with both greater psychological closeness and a higher proportion of participants meeting the criterion for early self-initiated engagement. Spatial and social cues may contribute to the perceived personal relevance of the initial experience, potentially encouraging subsequent voluntary engagement with the performers or related content.

## 4.4 Evaluating Immersive Video Beyond Exposure

Research on Presence in immersive media has often focused on the subjective experience of "being there" during exposure (Slater, 2009; Cummings & Bailenson, 2016). Toida et al. (2026) extended this perspective by reporting that filming-distance conditions associated with different levels of Presence also produced different increases in psychological closeness to performers depicted in immersive video. The present study further introduces temporal and behavioural dimensions, examining not only what is experienced during exposure but also what persists after the experience and what subsequently occurs.

The present findings indicate that the evaluation of IV should extend beyond Presence during exposure. In addition to the magnitude of its immediate effects, attention should be directed towards what persists after the experience and what subsequent behaviours emerge. Thus, IV holds promise as a medium that elicits stronger immediate psychological responses than conventional 2D video, some of which persist beyond exposure, and may encourage subsequent self-initiated engagement, including more active or socially oriented behaviours. Future longitudinal studies that assess Presence, emotional arousal, psychological closeness, subjective lasting impressions, and objective behavioural logs within the same participants could more directly elucidate the temporal pathway from the initial experience to psychological persistence and subsequent self-initiated behaviour.

### 4.5 Limitations

Several limitations of the present study should be acknowledged.

First, the sample size was smaller than initially planned, limiting statistical power, particularly for between-group comparisons at later follow-up time points and for behavioural outcomes. In addition, the IV group experienced the video through an HMD in a controlled experimental setting, whereas the 2D group viewed it online using their own devices. Thus, the observed differences reflect the overall viewing experience rather than video format alone. Prior awareness of the target group also differed descriptively between conditions and was not adjusted for in the behavioural analyses.

Second, self-initiated behaviour was assessed through self-report, and the behavioural analyses were restricted to participants who completed both scheduled follow-up assessments. The classification of early engagement also incorporated an operational criterion based on cumulative contact days. Furthermore, the 5A framework was used as a descriptive taxonomy rather than a validated model of sequential behavioural change. In particular, A5 included a composite social-media response item and therefore did not necessarily indicate interpersonal recommendation.

Third, the stimulus was limited to a live performance by a Japanese idol group, and the generalisability of the findings to other content remains to be established. Because Presence was not directly measured in the present study, its relationship with the persistence of psychological closeness and subsequent self-initiated behaviour remains an important direction for future research.

## 5 CONCLUSION

This study examined the persistence of psychological closeness following Immersive Video exposure and subsequent self-initiated behaviour over a two-week period. Psychological closeness was significantly higher in the IV group than in the 2D group immediately after exposure, and this advantage remained statistically significant through approximately seven days. Although the between-group difference was no longer statistically significant at 14 days, psychological closeness in the IV group remained elevated above its pre-exposure level. Lasting-impression ratings were also significantly higher in the IV group throughout the post-exposure assessments. Furthermore, early self-initiated engagement was identified in 17 of 37 participants in the IV group, compared with 5 of 38 in the 2D group; the proportion was approximately 3.5 times as high in the IV group. A higher proportion of IV participants also reported active or socially oriented behaviours during follow-up, although statistical evidence for this difference was less conclusive.

These findings suggest that the effects of IV are not necessarily confined to the immediate viewing experience. Under the viewing conditions examined in this study, IV was associated with greater psychological closeness to the target group than conventional 2D video, with some of this advantage persisting beyond exposure. A higher proportion of participants in the IV group also met the criterion for early self-initiated engagement. Thus, evaluating immersive media requires attention not only to the immediate experience but also to the psychological and behavioural responses that follow it.

## ACKNOWLEDGEMENTS

The idol group STU48 performed the song and choreography featured in the experimental video stimulus. WATANABE Toru and IWAMI Shota (Concent, Inc.) assisted with the filming of the immersive video. NATSUKA Masato (CRAZY TV Creative, Inc.) assisted with the audio recording of the immersive video. The implementation of the immersive video for Apple Vision Pro was carried out by MATSUBARA Tatsuro (MESON, Inc.).

## CONTRIBUTIONS

KT, SF, SMi, and SMe contributed to the formulation of the research hypotheses, discussed the experimental design, and contributed to the interpretation and discussion of the findings. KT, SF, SMi, YK, and SMe conducted the experiment. KT designed the experiment, analysed the data, and wrote the manuscript. KT, SF, SMi, and SMe reviewed the manuscript. SMe was responsible for project administration, including the oversight and coordination of the broader Immersive Video research project. All authors approved the final version of the manuscript.

## CONFLICTS OF INTEREST

This study was conducted as a joint research project between MESON, Inc. and Hakuhodo DY Holdings Inc. MESON, Inc. was involved in the production and implementation of the immersive video stimulus, while Hakuhodo DY Holdings Inc. operates in the marketing and communications sector. Both organisations may have professional or commercial interests in applications of immersive media. The authors declare no additional competing interests.